\documentstyle[11pt,newpasp,twoside,epsf]{article}
\def\nii{[N~{\sc ii}]}
\def\sii{[S~{\sc ii}]}
\def\oii{[O~{\sc ii}]}

\def\oiii{[O~{\sc iii}]}
\def\ha{H$\alpha$}

\def\edcomment#1{\iffalse\marginpar{\raggedright\sl#1\/}\else\relax\fi}
\reversemarginpar

\begin{document}
\title{New Results on R~Aquarii}
\author{Denise R. Gon\c calves, Antonio Mampaso, and Silvana Navarro}
\affil{Instituto de Astrof\'\i sica de Canarias,
       E-38205  Tenerife, Spain}
\author{Romano L. M. Corradi}
\affil{Isaac Newton Group of Telescopes, Apartado de Correos 321,
E38700 Santa Cruz de la Palma, Spain}

\begin{abstract}

The first results on new optical data for R~Aquarii\footnote{Based on
observations obtained with the 2.5m INT, the 2.5m NOT and the 3.5m NTT.}  are
presented.  The morphology and kinematics of the nebula, based on data obtained
with the NTT from 1991 to 2000, are discussed.  Physical parameters of the outer
nebula and the knotty jet are derived using spectra obtained with the INT in
2001.  From the analysis of all these data we propose that the spectacular
knotty inner structure of R~Aqr could result from the interaction of a highly
collimated pulsating young jet with the older hourglass inner nebula.

\end{abstract}

\section{The R~Aqr System}

Since the work of Solf \& Ulrich (1985), it has been well known that the
large-scale optical structure of R~Aqr consists of two binary (hourglass-like)
shells formed by two successive explosions of the system, and that these shells
share the same major axis.  The inclination of the polar axis with respect to
the line of sight is $\sim$ $70\deg$ (Hollis et al.~1999).  The expansion of
both shells in the polar direction is about six times faster than that in the
equatorial direction, their polar expansion being 32~km~s$^{-1}$ and
55~km~s$^{-1}$, implying kinematical ages of 185~yr and 640~yr, for the inner
and outer shells, respectively (Solf \& Ulrich 1985).  On smaller scales, R~Aqr
has a string of knots whose first detection occurred in the late 1970s (the NE
jet:  knots A, B and D in Fig.  3) and 1980s (the SW jet:  knot A$'$ in Fig.  3;
see Paresce, Burrows, \& Horne 1988).  Hollis et al.~(1991) derived the
kinematical age of the outermost NE knot as being around 90~yr, implying that
this jet is younger than the inner large-scale shell.  Although being around
90~yr old, the jet was not observed before 1977 (NE knots, Wallerstein \&
Greenstein 1980; Herbig 1980) nor before 1988 (SW knot, Hollis, Wagner \&
Oliversen 1990).  Its sudden detection, with some brightening enhancement,
suggests its impact with the environments (the inner shell).

\begin{figure}[!t]
\plotone{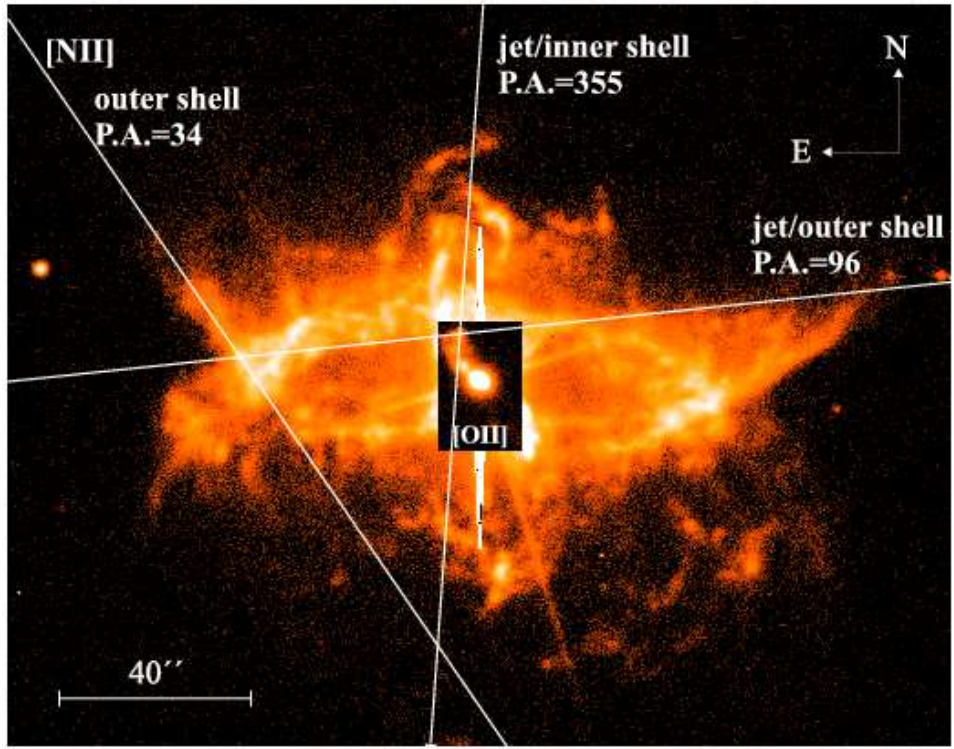}
\caption{The \nii\ image of R~Aqr large-scale shells, with the \oii\ jet 
image in the center. The vertical bright ``bar'' is caused by the 
saturation in the central region. The white lines show schematically the slit 
positions 
for the INT + IDS spectra (see text).}
\end{figure}

\section{Physical Parameters and Excitation of Outer Shell and Inner Jet}

We derived the physical parameters and excitation mechanisms for the different
features of R~Aqr, based on the 3.11 \AA/pix, $0.7\arcsec$/pix long-slit
spectra, obtained with the INT + IDS in August, 2001.  Fig.  1 shows the
positions of our three slits superposed on the image of the large and
small-scale features of R~Aqr.  We choose these three PAs in order to study the
properties of the jet's knots and inner shell ($3.8\arcsec$ E, $8.4\arcsec$ N,
P.A.  = $355\deg$), those of the outermost jet's knot and outer shell
($3.8\arcsec$ E, $8.4\arcsec$ N, P.A.  = $96\deg$) and of the brightest regions
of the outer shell ($42.5\arcsec$ E, $2.1\arcsec$ N, P.A.= $34\deg$).

Electron temperatures and densities, $N_e$\sii, were estimated at many positions
along the three slits.  Portions of the outer shells have, on average,
$T_e$\nii\ $\cong 1.4 \times 10^4$~K ($T_e$\oiii\ $\cong 1.8 \times 10^4$~K).
The jet's knots have $T_e$\nii $\cong 1.3 \times 10^4$~K and $T_e$\oiii $\cong
1.9 \times 10^4$~K, implying very similar $T_e$ for the outer shell and inner
jets.  Densities, on the other hand, vary by large amounts along each slit.  The
NE portion of the outer shell, crossed by two slits, has a $N_e$\sii\ $\cong
230$~cm$^{-3}$.  The rougly opposite position of the outer shell, its NW side,
shows approximately the same density, $\sim 200$~cm$^{-3}$.  At the position of
Knot~D, in which the P.A.= $96\deg$ as well as P.A.= $355\deg$ are centred,
densities are higher than those that can be safely determined by the \sii\
lines, $N_e$\sii\ $>10^4$~cm$^{-3}$.  Finally, electron densities of the inner
knots are around 600~cm$^{-3}$, and that of the inner shell cut by our slit at
the Northern as well at Southern sides of the system (P.A.  = $355\deg$) is
$\sim$~620~cm$^{-3}$.  At the positions where other estimations are available,
they are in general agreement with ours (Solf \& Ulrich 1985; Kafatos,
Michalitsianos, \& Hollis 1986; Hollis et al.~1991; Meier \& Kafatos 1995).

\begin{figure} 
\plotfiddle{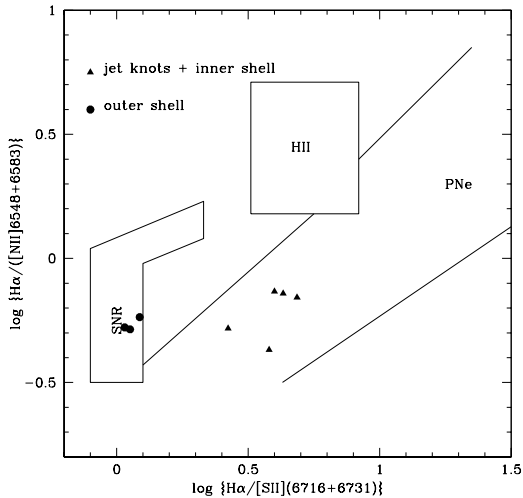}{6truecm}{0}{35}{35}{-90}{-60}
\caption{Diagnostic diagram with \oiii/ha \ \ versus \sii/\ha\ showing  
different features of R~Aqr (from Cant\'o 1981).}
\end{figure}

The high $T_e$\oiii\ ($\sim~1.8 \times 10^4$~K) measured for the system is
indicative of mild shock excitation.  With Fig.  2 we investigate the excitation
of many regions in R~Aqr.  We see that those regions which are mainly excited by
shocks are part of the outer shell (circles).  On the other hand, all the other
features (triangles) lie in the zone of the diagram where photoionization by the
central star is the main excitation process (the PNe zone).  Note, however, that
PNe used to define the PNe ``zone'' are not spatially resolved, at variance with
our points for R~Aqr in Fig.  2.  Because of that it is possible that some shock
excitation could be contaminating features placed in this zone of the diagram.

\subsection{Morphological and Kinematical Jet Evolution}

From the images in Fig.  3 (1991 July, with the NTT + EMMI, $0.35\arcsec/$pix;
and 1997 July with the NOT + ALFOSC; $0.19\arcsec$/pix), and previous works, the
morphological evolution of the jet is the following:  Knot~B was the brightest
one up to 1985 (Paresce et al.~1988); Knot~B and Knot~D were as bright as Knot~A
in 1986 (Solf 1992); Knot~D was hardly brighter than Knot~B in 1991 (Fig.  3);
and finally, Knot~A and Knot~A$'$ were the brightest in 1997 (Fig.  3).  We also
note that the knotty features evolved from round and compact knots in 1991 to
more elongated and diffuse ones in 1997.  From the morphological changes of Fig.
3, and references cited above, it is evident that the R~Aqr jet evolves over
timescales smaller than 5~yr.  However, caution should be taken when comparing
images obtained in different epochs and with different instrumentations, filters
and seeings.

From the high resolution spectra (1999 January, see Navarro et al., this volume,
p.  000), one of them crossing the Knot~D, we note that in 1999 Knot~D was no
longer the brightest knot, since it was not detected in our spectrum, at
variance with knots A and A$'$.  Therefore, if present in 1999, Knot~D came to
be less bright than the innermost ones Knot~A (NE jet) and Knot~A$'$ (SW jet).

\begin{figure} 
\plotfiddle{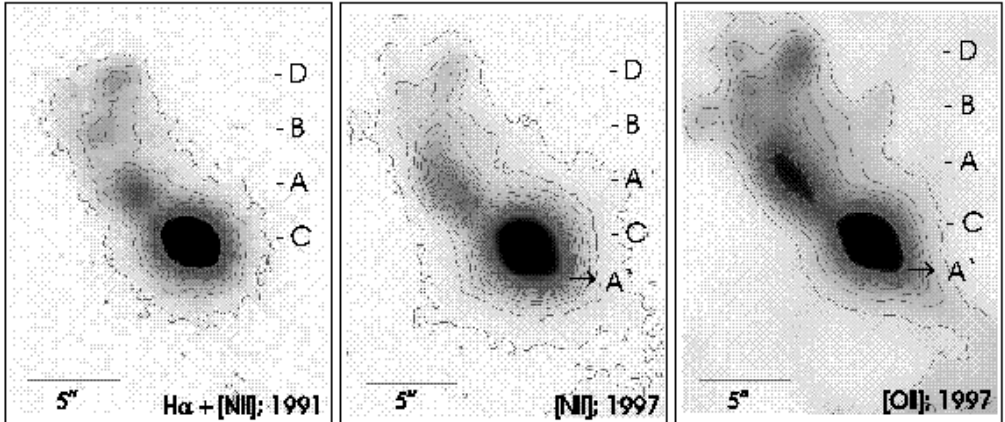}{4.5truecm}{0}{66}{66}{-200}{-195}
\caption{Images of the R~Aqr inner jet. {\itshape Left to right;\/} The 
broad \ha, and narrow \nii\ and \oii\ filters images in 1991, 1997, 
and 1997, respectively. Note the horizontal marks on the right side of
each panel that define the position of the Knots A, B and D (NE jet components);
C (the central source continuum) and  A$'$ (the SW jet).}
\end{figure}

\subsection{On the Controversial Origin of the Knotty Jet}

Solf \& Ulrich (1985) and Solf (1992) first suggested that the string of knots
(A to D) is formed by fossil condensations of the inner (190~yr) shell, which
are being illuminated by the impact of a well collimated jet.  However, Hollis
and coworkers (Hollis et al.  1991, 1997, 1999) argumented that such a group of
knots are bright clumps of a highly collimated jet that interacts with the inner
nebulosity.  In another words, these two approaches differ in that the former
sees the knots as part of the inner shell and the latter as part of the jet.
Finally, since Hollis et al.  (1991) determined the age of the outermost jet's
knot (Knot~D) as being $\sim$~90~yr, it became clear that knots cannot be
inhomogeneities of the inner shell, which is considerably older than Knot~D (see
Section 1).

The present results are much more in agreement with the idea that the string of
knots in the inner regions of R~Aqr is part of a knotty jet, because of the
brightness' evolution of the knots (discussed in the previous section),
indicative of a precessing jet with pulsation.  Putting together the latter, the
new results on the kinematics of the knotty jet (see also Navarro et al, this
volume, p.  000) and the jet's age (Hollis et al.  1991), we propose that the
knotty inner structure of R~Aqr is the result of the interaction of a highly
collimated pulsating young jet with the older hourglass inner nebula.  To check
further this idea a more robust comparison of images (probably compiling data
from many authors in order to achieve a well covered set of filters and epochs)
is desirable, as well as more deeply investigating if the emission of the knots
is partially excited by shocks, which would prove their interaction with the
surrounds.

\section*{Discussion}
\noindent{\itshape Balick:\/} In R Aqr we see a precessing string of knots close
to the nucleus. In He 2-104 (which is about ten times more distant), we see an
hourglass nebula in the inner part of the object. Do you believe that there is an
evolutionary connection? Can the precessing string of outflowing knots eventually
form a larger hourglass, like that in He 2-104 or MyCa18?

\smallskip

\noindent{\itshape Gon\c calves:\/} Considering that we have precession in the very
inner R Aqr jet, we might think about an evolution in this system that  would result
in structures like the hourglass shells of, for instance, He 2-104. However, if we 
compare the R Aqr jet with that of CH Cyg, we would say that such an evolution
is not that clear. In CH Cyg, the radio jet observed in 1985 (Taylor et al.~1986)
 disappeared about fifteen years later (Corradi et al.~2001); in the optical emission,
 its remnant is what we see now as an optical nebulosity, which appears in the
 PV diagrams as having a more or less hourglass structure. But note that the 
 latter is much more of a speculation than a clear result based on the data.

\smallskip

\noindent{\itshape Viotti:\/} Also in connection with the possible origin of the 
soft X-ray emission of R Aqr detected by {\itshape EXOSAT}, it would be desirable
to investigate the ionization in different parts of the nebula, for instance by
looking at the He$^{++}$/He$^{+}$ emission lines.

\smallskip

\noindent{\itshape Gon\c calves:\/} Only the spectra of the jet present these lines
and they are fainter than, for instance, the [O~{\sc iii}], [S~{\sc ii}],
[N~{\sc ii}] lines, which I have used to derive the main excitation mechanism
of each region (Phillips \& Cuesta 1999; Cant\'o 1981; etc.).


\end{document}